\documentclass[twocolumn,showpacs,superscriptaddress,preprintnumbers,amsmath,amssymb,prl]{revtex4}

\usepackage{graphicx}
\usepackage{dcolumn}
\usepackage{bm}

\begin{document} 
\newcommand{\bc}{\begin{center}}
\newcommand{\ec}{\end{center}}
\newcommand{\nn}{\nonumber}
\title{Tomonaga-Luttinger Liquid Renormalized by a Single Impurity}
\author{Kenji Kamide}
\affiliation{Department of Physics, Waseda University, 3-4-1 Okubo, Shinjuku, Tokyo 169-8555, Japan}
\author{Takashi Kimura}
\affiliation{Department of Information Sciences, Kanagawa University, 2946 Tsuchiya, Hiratsuka, Kanagawa 259-1293, Japan}
\author{Susumu Kurihara}
\affiliation{Department of Physics, Waseda University, 3-4-1 Okubo, Shinjuku, Tokyo 169-8555, Japan}
\email{kamide@kh.phys.waseda.ac.jp}
\date{\today}

\begin{abstract}
Effects of a single impurity potential on bulk properties of a spinful Tomonaga-Luttinger (TL) liquid is studied.
A boundary bosonization technique is developed to include the impurity potential of {\it arbitrary} strength~$V$.
Our new bosonization formula for fermion field $\psi$ smoothly connects the two existing expressions in  the strong ($V = \infty$) and the weak ($V = 0$) impurity limits.
With use of the formula, we found the TL parameters determined from the long-distance correlation functions are renormalized due to the partial transmission through the impurity potential.

\end{abstract} 
\pacs{71.10.Pm, 71.30.+h, 72.10.-d, 73.23.-b, 73.40.Gk, 73.63.-b}
\maketitle 

Interactions between particles introduce collective features into many-body systems.
The dramatic many-body effects are reported in a variety of one-dimensional (1D) systems such as quantum wires~\cite{Yacoby}, carbon nanotubes~\cite{Ishii}, neutral atomic gases of fermions~\cite{Moritz} and bosons~\cite{Thilo,Bloch} in 1D optical lattices.
A single-particle excitation loses its weight by exciting infinite numbers of particle-hole pairs. 
Low energy physics of interacting 1D particles are described by the collective excitations of density fluctuations: Tomonaga-Luttinger (TL) liquid~\cite{Tomonaga,Luttinger,Giamarchi}. {\it Only} the collective features emerge in the lowest energy scales for homogeneous systems. 

Once inhomogeneity is introduced to the system by a localized potential $V(x) \sim V\delta(x)$, such as an impurity, the single-particle excitations come into play due to their backscattering, and strongly affect the lowest collective excitations.
They cause the zero-bias anomaly in the 1D electron transport~\cite{West,Bachtold}.
Effects of a single impurity in a TL liquid have been intensively studied in the context of the scaling of impurity potential~\cite{KF,FN}.
Roughly speaking, for a spin independent system, the electron transmission probability scales toward 1 or 0 as lowering temperature, depending on whether the interparticle interaction is attractive or repulsive. 
This is a prediction from the perturbative renormalization group (RG) analysis near the two fixed point values, $V=0$ or $\infty$.
The scaling flows at arbitrary $V$ have not been accessible with some exceptions, the model for which the exact solution is available at special interaction parameters~\cite{Fendley}, or for weak interaction limits near Fermi-liquid fixed points~\cite{Glazman,Stauber}. 
Recently, the numerical assessment was done for intermediate region by a path-integral Monte Carlo method~\cite{Hamamoto}. However, their result shows in some part non-trivial behaviors in the scaling flows of the zero-bias conductance. 
Thus, the effect of a single impurity of arbitrary strength is still unclear after the pioneering works of 15 years ago.

Theoretical difficulties in treating arbitrary $V$ are concentrated on the point that both the long-wavelength and short-wavelength physics are of equal importance.
This bears a resemblance to the Kondo effect~\cite{Kondo} and the Fermi-edge singularity in X-ray absorption spectrum~\cite{Mahan,Shotte&Ogawa}.
The difference from them is in the point that the local degrees of freedom and the collective degrees of freedom originate from the same conduction electrons.
Hence, a unified theory describing the single-particle field and long-wavelength collective TL field is needed. The aim of this paper is to construct the bosonization formula which gives the relation between the single-particle phase shift $\theta_{\rm F}$ and the bulk excitations of the collective mode~\cite{kamide1,Aristov,kamide2}.
The power law dependencies of various correlation functions are calculated with use of the obtained expression.
We show explicitly the bulk excitations suffers a significant change due to the presence of a single impurity potential.
The zero bias conductance is also determined for a given phase shift.
We set $\hbar=k_{\rm B}=1$, the inverse temperature $\beta=1/T$, and the system size $L$.

The Hamiltonian under consideration consists of three parts $H=H_0 +H_{\rm V} + H_{\rm int}$,   
\begin{eqnarray}
&&H_0=\int {\rm d} x ~\psi^{\dagger}_\sigma (x) \left( p_x^2/2m-\varepsilon_{\rm F} 
\right) \psi_\sigma (x) ,\\
&&H_{\rm V}= V \int {\rm d} x ~\delta(x) \psi_\sigma^{\dagger} (x)   \psi_\sigma (x) ,\\
&&H_{\rm int}=\int {\rm d} x \int {\rm d} x'  ~n_\sigma(x)  U_{\sigma , \sigma '}(x-x') n_{\sigma'}(x'), \label{eq:interaction-model}
\end{eqnarray}
where $\psi_\sigma$ and $n_{\sigma}=\psi_\sigma^{\dagger}\psi_{\sigma}$ are the electron field and density of spin $\sigma=\pm 1=\uparrow,\downarrow$. $H_0$, $H_{\rm V}$, and $H_{\rm int}$ represent, respectively, the kinetic energy, an impurity potential, and the two-body interactions. 

\begin{figure}[tbp]
\includegraphics[width=8.5cm]{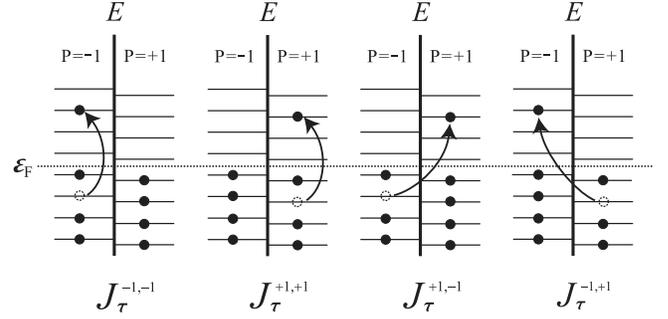}
\vspace{0mm}
\caption{\label{fig:distribution} Four types of the density operator $J_{\tau}^{\rm P,P'}$ represent a pair excitation of a particle with the parity $\rm P=\pm1$ and a hole with the parity ${\rm P'}=\pm 1$. In the each diagrams, energy levels of the even (odd) parity states are shown in the right (left) side of the vertical axis (energy) where the particles (holes) are represented by filled (open dotted) circles.}
\end{figure}

Before treating the spinful model, let's first derive a bosonization formula for a spinless model. 
We start with the one-body scattering problem
\begin{eqnarray}
\left( - \partial_x^2/2m+V \delta(x) \right) \phi(x) = \varepsilon_k \phi (x).  \label{onebody}
\end{eqnarray}
The solutions of even ($P=+1$) and odd ($P=-1$) parity are: 
$\phi_{{\rm P}=1,k}(x)=\sqrt{2/L} \cos \left( |kx|-\theta_{\rm F} \right)$ and $\phi_{{\rm P}=-1,k}(x) =\sqrt{2/L}\sin \left(kx \right)$.
The momentum $k$ is defined in a semi-infinite space $k>0$, and the scattering phase shift $\theta_{\rm F}=\tan ^{-1}(mV/k_{\rm F})$ is approximated at the Fermi level. 
The fermion field is given by $\psi(x)=\sum_{{\rm P}, k>0}\phi_{{\rm P},k}(x) c_{{\rm P},k}$, where $c_{{\rm P},k}~(c_{{\rm P},k}^{\dagger})$ is annihilation (creation) operator of a state $({\rm P},k)$.
In treating the two-body interaction in bosonization technique, it is convenient to use the decomposition
\begin{eqnarray}
\psi(x) \sim \tilde{\psi}_{+}(x)e^{ik_{\rm F}x}+\tilde{\psi}_{-}(x)e^{-ik_{\rm F}x}.  \label{eq:fermionfield0}
\end{eqnarray}
This procedure is brought forward easily for the homogeneous case.
In this case, however, one must keep in mind that these two chiral fields depend on each other.
To fulfill the constraint, we introduce a fermion operator $\tilde{c}_{{\rm P},k}$, which is defined in a full momentum space ($-\infty<k<\infty$) and satisfies $\tilde{c}_{{\rm P} ,k}=\tilde{c}_{{\rm P} ,-k}(=c_{{\rm P} ,k})$. In addition, we keep the anti-commutation relation $\{\tilde{c}_{{\rm P} ,k}^{\dagger},\tilde{c}_{{\rm P}' ,k'}\}=\delta_{\rm P,P'}\delta_{k,k'}$. 
The chiral field $\tilde{\psi}_\tau$ ($\tau =\pm $) consists of two field
\begin{eqnarray}
\tilde{\psi}_{\tau}(x) &=& e^{-i \tau {\rm sgn}(x) \theta_{\rm F}} \tilde{\psi}_{+1,\tau}(x) + e^{-i\tau \pi/2} \tilde{\psi}_{-1,\tau}(x),
\end{eqnarray}
where each of them comes from the even or odd functions,
\begin{eqnarray}
\tilde{\psi}_{{\rm P},\tau}(x) &=&\frac{1}{\sqrt{2L}} {\sum}_{k >0} e^{i \tau (k-k_{\rm F})x} \tilde{c}_{{\rm P},\tau k} .  
\end{eqnarray}
The slowly varying density $J_{\tau} (x) \equiv \tilde{\psi}^{\dagger}_\tau (x) \tilde{\psi}_\tau (x)$ is given by $J_{\tau}^{\rm P,P'}(x) \equiv \tilde{\psi}^{\dagger}_{{\rm P},\tau}(x)\tilde{\psi}_{{\rm P'},\tau}(x)$ as
\begin{eqnarray}
J_{\tau}(x) &=&  {\sum}_{\rm P, P'} e^{- i \tau \theta_{\rm P,P'}(x)} J_{\tau}^{\rm P, P'}(x) \nonumber \\
&=& \frac{1}{2L} {\sum}_{{\rm P,P'},q}  e^{iqx - i \tau \theta_{\rm P,P'}(x)} J_{\tau}^{\rm P, P'}(q) . \label{eq:density1}
\end{eqnarray}
The four density operators corresponds to the four kinds of particle-hole excitations characterized by the parities of a particle and a hole wavefunction (see Fig.~\ref{fig:distribution}).
The scattering phase shift enters the above expression as
\begin{eqnarray}
&& \theta_{\rm P, P'}(x)=\delta_{\rm P,-P'} \times {\rm P} \times  (  {\rm sgn}(x)  \theta_{\rm F} -\pi/2 ). \label{eq:phaseshift}
\end{eqnarray}
It is worth noting that the phase shift appears only in the density operators with odd parity ${\rm PP'}=-1$.
It is naturally understood from the fact that in order to excite the density wave of odd parity, a relative number of fermions between the right and the left of the impurity must change, i.e. the particles must tunnel through the impurity potential.
It is convenient to express the four kinds of bosonic density operators in terms of pseudo spin defined by $\tilde{\zeta}_{\tau}(x)=(\tilde{\psi}_{+1,\tau}(x),\tilde{\psi}_{-1,\tau}(x))^{t}$. The pseudo spin density is defined by ($i=0,1,2,3$)
 \begin{eqnarray}
J_{\tau}^{i}(x)= \frac{1}{\sqrt{2}}  \tilde{\zeta}_{\tau}^{\dagger}(x) \gamma_{i}     \tilde{\zeta}_{\tau}(x) ,
\end{eqnarray}
where $\gamma_0$ is a $2 \times 2$ unit matrix, and $(\gamma_1,\gamma_2,\gamma_3)$ are Pauli matrices in the pseudo spin space. The pseudo spin densities satisfy the commutation relation, 
\begin{eqnarray}
\left[  J_\tau^{i}(q),J_\tau^{j}(-q') \right]=\delta_{i,j}\delta_{q,q'} (\tau qL/2\pi). \label{commutation1}
\end{eqnarray}
We rewrite $H_0+H_{\rm V}$ in bilinear form of them by linearising the spectrum around the Fermi level as 
\begin{eqnarray}
H_0 + H_{\rm V}&=& \frac{\pi v_{\rm F}}{L}\sum_{i=0,1,2,3} \sum_{\tau, q} J_{\tau}^{i}(q)J_{\tau}^{i}(-q). \label{H0}
\end{eqnarray}
Next, we will consider interactions between particles.
We assume that the relevant physics originating from the particle correlations can be captured with a $(g_2.g_4)$-model~\cite{Solyom} which takes into account only the forward scatterings and neglects the backward scattering $g_1$.
This approximation is valid quantitatively as far as the Fourier components of the interaction potential in Eq.~(\ref{eq:interaction-model}) satisfy $\tilde{U}_{\sigma,\sigma'}(q \sim 0)\gg \tilde{U}_{\sigma,\sigma'}(q \sim 2k_F)$ as in the case of long-range interaction.
Then, the interaction Hamiltonian can be written 
\begin{eqnarray}
 H_{\rm int}&=&{\sum}_{\tau} \int {\rm d} x  \left[ g_4 J_{\tau}(x) J_{\tau}(x) +g_2 J_{\tau}(x)J_{-\tau}(x) \right] \nonumber \\
&=& \frac{1}{2L}{\sum}_{\tau, q}
\left[ g_4 J_{\tau}^{0}(q)J_{\tau}^{0}(-q)+g_2 J_{\tau}^{0}(q)J_{-\tau}^{0}(-q)
\right. \qquad \nonumber \\
&+& \sin^2(\theta_{\rm F})
\left(g_4 J_{\tau}^{1}(q)J_{\tau}^{1}(-q)+g_2 J_{\tau}^{1}(q)J_{-\tau}^{1}(-q)
\right)  \\
&+&\left.  \cos^2(\theta_{\rm F}) \left( g_4 J_{\tau}^{2}(q)J_{\tau}^{2}(-q)-g_2 J_{\tau}^{2}(q)J_{-\tau}^{2}(-q) \right)\right]. \nonumber  \label{Hint}
\end{eqnarray}
In order to get the second equation, we kept only the terms nonvanishing after the integration over $x$ in the r.h.s. of the first equation. The nonvanishing terms satisfy  the parity conservation $({\rm P_1 P_2 P_3 P_4}=+1)$. 
According to the commutation relation in Eq.~(\ref{commutation1}), the bosonic ladder operators are introduced as
\begin{eqnarray}
J_{\tau}^{i}(q)=| qL/2 \pi |^{1/2} \left( 
\theta(\tau q) b_{i}(q) + \theta(-\tau q) b_{i}^{\dagger}(-q)
\right).
\end{eqnarray}
By introducing a Bogoliubov transformation
\begin{eqnarray}
\left(
\begin{array}{c}
b_i(q) \\
b_i^{\dagger}(-q)
\end{array}
\right)
=
\left(
\begin{array}{cc}
\cosh(\phi_i) &\sinh(\phi_i) \\
\sinh(\phi_i) &\cosh(\phi_i)
\end{array} 
\right)
\left(
\begin{array}{c}
\beta_i(q) \\
\beta_i^{\dagger}(-q)
\end{array}
\right),
\end{eqnarray}
the total Hamiltonian is diagonalized as $H=\sum_{q,i} v_{i}|q|(\beta^{\dagger}_{i}(q)\beta_{i}(q)+\frac{1}{2})$.
The rotation angles depend on dimensionless coupling parameters $\tilde{g}_{2(4)}=g_{2(4)}/2\pi v_{\rm F}$, 
\begin{eqnarray}
\tanh(2 \phi_0)&=&\frac{-\tilde{g}_2}{1 + \tilde{g}_4} ,~ 
\tanh(2 \phi_1)=\frac{-\tilde{g}_2 \sin^{2}(\theta_{\rm F})}{1 + \tilde{g}_4 \sin^{2}(\theta_{\rm F}) }, \nonumber  \\
\tanh(2 \phi_2)&=&\frac{\tilde{g}_2 \cos^{2}(\theta_{\rm F})}{1 + \tilde{g}_4 \cos^{2}(\theta_{\rm F}) } , ~
\tanh(2 \phi_3)= 0. \label{TLparameters1}
\end{eqnarray}
The four-component TL liquid is characterized by four TL parameters $K_i=e^{2\phi_i}$ ($i=0,1,2,3$).
The parameter for the charge $K_0$ is exactly the TL parameter in a homogeneous system. For clarity, we shall call $K(=K_0)$ {\it a bare TL parameter}.
We now reach our main result, the boson representation of the fermion field
\begin{eqnarray}
\tilde{\psi}_{\tau}(x) &\sim& \sqrt{k_{\rm F}/\pi}~e^{{\rm i} \tau \chi_{\tau}(x)}, \\
 \chi_{\tau}(x)&=&\frac{\pi }{L} \sum_{q,{\rm P,P'}}
\frac{e^{{\rm i}qx-{\rm i}\tau \theta_{\rm P,P'}(x)  } }{iq}J_{\tau}^{\rm P,P'}(q) \nn \\
&=& 
\frac{1}{i}\sum_{q>0} \sqrt{\frac{\pi}{qL}} \Big[ e^{{\rm i} \tau qx}\left(
\tau b_0(q) + \tau {\rm sgn}(x)\sin(\theta_{\rm F}) b_1(q) \right. \nn \\
&&
\left. - \cos(\theta_{\rm F})b_2(q)
\right) -{\rm h.  c.~} \Big]. \label{formula}
\end{eqnarray}
To get this expression, we put the aforementioned constraints $\tilde{c}_{{\rm P},k}=\tilde{c}_{{\rm P},-k}$ i.e. $b_i(q)=b_i (-q)$.
It is easily checked that this formula recovers the known expression in the weak and strong barrier limits by putting $\theta_{\rm F}=0$ or $\pi/2$.
For example, in the strong barrier limit~\cite{Eggert}, we have 
\begin{eqnarray}
\chi_{\tau}(x)&=&\tau \sum_{q>0}  \sqrt{\frac{2\pi}{qL}} \left( \sqrt{1/K} \cos{qx}+i\tau \sqrt{K} \sin{qx} \right) \nn \\
&& \qquad \quad \times \beta'( {\rm sgn}(x) q) +{\rm h.c.} , 
\end{eqnarray} 
where $\beta'(\pm q)=\left(\beta_0(q)\pm\beta_1(q)\right)/(i \sqrt{2})$.
With our formula, we clearly see the operators for $x>0$ and $x<0$ are separable for $\theta_F =\pi/2$.
In this case, the TL liquid is regarded as completely disconnected two pieces.

The extension of these results to a spinful model is straightforward when the potential barrier is independent of spin $V_\uparrow=V_\downarrow=V$, since charge~($c$) and spin~($s$) variables are completely decoupled in the TL Hamiltonian. It is simply done by adding real spin variables $\sigma=\pm 1=\uparrow,\downarrow$. The system is described as an eight-component TL liquid with the Hamiltonian $(H)_q=\sum_{i=0-3,j=c,s} v_{j,i}|q|(\beta^{\dagger}_{j,i}(q)\beta_{j,i}(q)+\frac{1}{2})$. Following the conventions~\cite{Solyom}, we denote the dimensionless coupling constant for the charge (spin) $\tilde{g}_{n}^{c(s)}=(\tilde{g}_n^\parallel \pm \tilde{g}_n^\perp)/2$ for $n=2,4$. TL parameters are defined $K_{c (s),i}=e^{2 \phi_{c (s),i}}$ where $\phi_{c(s),i}$ is given by putting $\phi_{i} \to \phi_{c (s),i}$ and $\tilde{g}_{n} \to \tilde{g}_{n}^{c(s)}$ in Eq.~(\ref{TLparameters1}). In the same way, the phase field $\chi_{\tau, \sigma}=(\chi_{\tau, c} +\sigma \chi_{\tau, s})/\sqrt{2}$ of the chiral fermion $\tilde{\psi}_{\tau,\sigma} \propto e^{i\tau \chi_{\tau, \sigma }}$ is obtained by putting $\chi_{\tau} \to \chi_{\tau,c(s)} $ and $b_{i}\to b_{c(s),i}$ in Eq.~(\ref{formula}).
In the case of the spin dependent potential barrier, a mixing of spin and charge degrees of freedom occurs through the difference in the scattering phase shifts.
In that case, however, one can solve the problem by introducing a rotation in the spin and charge space~\cite{Penc,kimura,kamide3}.

So far, we see the spinful TL liquid with a single impurity is described as eight-component TL liquid. This indicates the bulk theory is subject to change in the presence of a localized scatterer.
This change can be evaluated from the difference in the correlation functions in the bulk region $|x| \gg v_{\rm F} \beta $.
Similarly, the renormalization of TL parameters is also evaluated from the scaling dimensions of the long distance correlation functions~\cite{Giamarchi}. 
The correlation functions of charge density wave $O_{\rm cdw}=\tilde{\psi}^{\dagger}_{\tau,\sigma}\tilde{\psi}_{-\tau,\sigma}$, spin-density wave (sdw), singlet superconductivity (ss), and triplet superconductivity (ts) are calculated as $\langle O^{\dagger}(x)O(x')\rangle \propto \left|x-x' \right|^{-\kappa}$ for $|x|,|x'|\gg v_{\rm F}\beta$. The exponents are 
\begin{eqnarray}
\begin{array}{ll}
\kappa_{\rm cdw}=A_{c,+1}+A_{s,+1}, & \kappa_{\rm sdw}=A_{c,+1}+A_{s,-1}, \\
\kappa_{\rm ss}=A_{c,-1}+A_{s,+1}, & \kappa_{\rm ts}=A_{c,-1}+A_{s,-1},
\end{array} 
\label{exponents}
\end{eqnarray} 
where, for $j=c,s$ and $\nu=\pm 1$,
\begin{eqnarray}
2A_{j,\nu}=(K_{j,0})^{\nu }+\sin^2(\theta_{\rm F})(K_{j,1})^{\nu }+\cos^2(\theta_{\rm F})(K_{j,2})^{-\nu }. \nn 
\end{eqnarray}
By comparing Eq.~(\ref{exponents}) with those of homogeneous systems~\cite{Giamarchi}, the renormalized TL parameters~($K_{c}^{\ast},K_{s}^{\ast}$) of an effective two-component TL liquid must satisfy
\begin{eqnarray}
(K_{c}^{\ast})^{\nu }+(K_{s}^{\ast})^{\nu ' }=A_{c,\nu}+A_{s,\nu '}, \label{mapping}
\end{eqnarray}
for arbitrary combinations of $\nu,\nu'=\pm 1$.
\begin{figure}[btp]  
\begin{flushleft}
\includegraphics[height=4.5cm]{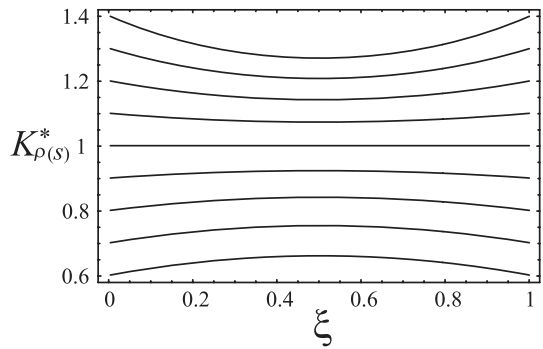} 
\end{flushleft}
\vspace{-4mm}
\caption{\label{fig:TLparameter} TL parameters $K_{c (s)}^{\ast}$ determined by Eq.(\ref{newparameter}) are plotted as function of transmission probability $\xi$ with for several values of the interaction parameter $\tilde{g}_{2}^{j}$ with $\tilde{g}_4^j=\tilde{g}_2^j$.}
\end{figure}
Generally, mapping the eight-component TL liquids to two-component model is impossible. However, a solution of a set of two equations derived from Eq.~(\ref{mapping}), 
\begin{eqnarray}
K_j^{\ast}=\sqrt{1+\left(\frac{A_{j,+1}-A_{j,-1}}{2}\right)^2}+\frac{A_{j,+1}-A_{j,-1}}{2}, \label{newparameter}
\end{eqnarray}
gives the reasonable value since it is exact at least to the linear order in the small coupling constants $\tilde{g}'s$, 
\begin{equation}
K^{\ast}_j = 1-\left( \left(\xi -\frac{1}{2}\right)^2+\frac{3}{4} \right)(\tilde{g}_2^{j}+\tilde{g}_4^{j}).
\end{equation} 
Thus, we accept $K^{\ast}_j$ in Eq.~(\ref{newparameter}) as the effective TL parameter.
$K^{\ast}_j$ is plotted as a function of $\xi$ in Fig.~\ref{fig:TLparameter}. 
The result indicates that {\it the TL parameter, which characterizes the bulk quantity of the lowest energy scale in a TL liquid, is renormalized due to the presence of a single impurity.}
Since our calculation starts from a scattering problem, all the single-particle states are under the effects of the single impurity potential and implicitly includes the Friedel oscillation~\cite{Glazman} which has the singular effects on the electronic correlations especially in one dimension.
It is worthwhile to note that $K^{\ast}_j$ approaches 1 for the intermediate value of $\xi$ in Fig.~\ref{fig:TLparameter}.
This indicates that the partial transmission (or the partial reflection) destabilizes the two-body interaction effectively than in the homogeneous system ($| K_{j,0}-1|\ge  |K^{\ast}_j (\xi)-1|$). This could be understood as follows.
The impurity scattering causes the phase mismatch between the density waves through Eq.(\ref{eq:phaseshift}).
Thus, there occurs a decrease in the overlap integral of the density waves due to the presence of the impurity scattering. This results in the weakening of the two-body interaction.
While the TL parameters depend on the strength of the impurity scatterings, the zero temperature phase diagram, which tells the most dominant correlation functions, remain the same as in a clean system~\cite{Solyom}; From Eq.~(\ref{exponents}), the phase boundaries are determined by $A_{c,+1}=A_{c,-1}$ and $A_{c,+1}=A_{c,-1}$, which is $K_{c,0}=K_{s,0}=1$.

Finally, we refer to the transport properties. 
The charge current is defined by $j_c(x,t)=\sum_{\tau} \partial_t \chi_{ \tau, c}(x,t)/2\pi$.
The zero bias conductance~$G$ is calculated with the use of Kubo formula and Eq.~(\ref{formula}).
Since a DC electric field which is symmetric $E(x)=E(-x)$ and concentrated around the impurity only couples with the current of a charge mode of $i=2$, we obtain
\begin{eqnarray}
 G &=&\frac{e^2}{\pi}K_{2,c}^{-1}\cos^2(\theta_{\rm F}) \nonumber \\
 &=&\frac{e^2}{\pi}\sqrt{\frac{1+(\tilde{g}_{4}^{c}-\tilde{g}_{2}^{c})\cos^2(\theta_{\rm F})}{1+(\tilde{g}_{4}^{c}+\tilde{g}_{2}^{c})\cos^2(\theta_{\rm F})}}\cos^2(\theta_{\rm F}). \label{linear conductance}
 \end{eqnarray}
This result shows resemblance to the Landauer formula for non-interacting quantum wire $G/G_0 =\cos^2(\theta_{\rm F})$ with difference of the prefactor.
However, it should be noted that the fermion phase shift given by Eq.~(\ref{onebody}) is exact only for noninteracting systems, since the phase shift, which is determined by the impurity potential $V$, should be modified by the electronic correlations.
RG equation of the phase shift may be evaluated from the correlation functions of boundary operators, such as the local density of states~\cite{kamide1}. 
One can reproduce the RG equation valid for arbitrary transmission but for the weak interaction~\cite{Glazman}. Further discussions on the renormalization of the transmission amplitude will be shown somewhere else.

In conclusion, we have studied the bulk properties of a TL liquid with a single impurity potential using our new bosonization method. Our theory relates a one-body scattering phase shift to the properties of the collective excitations. The new bosonization formula smoothly connects the existing formulae valid for $\theta_{\rm F}=0$ or $\pi/2$. TL parameters, which characterize the long distance correlation functions, are found to be renormalized due to the presence of a single impurity.
 
We acknowledge L. I. Glazman, K. A. Matveev, A. Furusaki, Y. Tsukada, N. Yokoshi, Y. Hamamoto, and D. Yamamoto for enlightening discussions. 
This work is supported by the Japan Society for the Promotion of Science, and Waseda University Grant for Special Research Projects.

\end{document}